\documentclass[
aps,%
10pt,%
final,%
notitlepage,%
oneside,%
twocolumn,%
nofootinbib,%
superscriptaddress,%
noshowpacs,%
centertags]%
{revtex4}
\usepackage{multirow}

\begin{document}

\title{GAS AND DUST IN THE BCD GALAXY VII ZW 403 (UGC 6456)}

\author{\firstname{O.V.}~\surname{Egorov}}
\email{morzedon@gmail.com}
\affiliation{Sternberg Astronomical Institute,
Universitetskiy pr. 13, Moscow, 119992 Russia}

\author{\firstname{T.A.}~\surname{Lozinskaya}}
\email{lozinsk36@mail.ru}
\affiliation{Sternberg Astronomical Institute,
Universitetskiy pr. 13, Moscow, 119992 Russia}

\accepted{by Astrophysical Bulletin}

\begin{abstract}Based on the results of the earlier spectroscopic
observations with the \mbox{6-m BTA} telescope of the SAO RAS we refine the metallicity
estimates of the complexes of ionized gas in the VII~Zw~403 galaxy. Infrared
observations from the Spitzer Space Telescope are used to search for a possible
correlation of the  mass fraction distribution of the polycyclic aromatic
hydrocarbons (PAH) with the distribution of ionized and neutral hydrogen, and with
the metallicity of gas in the HII regions of the galaxy.

\end{abstract}

\maketitle

\section{INTRODUCTION}

Detailed study of dwarf galaxies is important for understanding
the origin and evolution of galaxies. According to current ideas,
they play a role of building blocks, which are in their turn
responsible for forming more massive systems \cite{zhao,
kauffman}. The nearby blue compact dwarf galaxies (BCD) are of
particular interest if the distances to them allow detailed
studies of their structure, kinematics and emission spectra.

This paper continues a series of our investigations of ionized
gas complexes, associated with the latest burst of star formation  in the
BCD galaxy VII~Zw~403 (UGC~6456), presented  by Arkhipova et al.
\cite{our_2} and Lozinskaya et al. \cite{our_1}.

VII Zw 403 is one of the closest BCD galaxies ($D=4.5$\,Mpc), which has revealed
several episodes of star formation of different intensity \cite{lynds, shul_lad_a}.
According to Lynds et al. \cite{lynds}, the most powerful burst of star formation
occurred 600--800\,Myr ago, the stars of this generation dominate in the galaxy. The
age of the latest, fainter burst is about $4 - 10$~Myr. ~Luz  and \mbox{Thuan
\cite{luz_thuan}} have classified VII~Zw~403 as iE, the most numerous class of BCD
galaxies, which is characterized by an irregular bright star-forming region near the
center of an extended elliptical halo of old stars. Photometry of the stellar
population of VII~Zw~403   in the near IR confirmed that the region of recent
star formation is surrounded by an extended ``old'' stellar halo \cite{shul_lad_b}. The
latest burst of star formation encompassed the central region of the galaxy, spanning
about 1\,kpc in the direction of a giant cloud of neutral hydrogen with the
highest radiation density, surrounded by an extended \mbox{HI halo} sized $3.6'\times
2.9'$ ($4.7\times 3.8$\,kpc) \cite{thuan_04, ashley}. Several sites of this
latest star formation episode are observed: the youngest massive stars form compact OB
associations \mbox{¹~~1--6}  (for consistency, we use the nominations of
associations and HII regions \mbox{according to \cite{lynds}}).

The ionized gas is concentrated in the same central region; the
associations are linked with bright HII regions sized 80--150\,pc
surrounded by the faint diffuse  H$_\alpha$  emission
\cite{lynds, martin, thuan_87}. The
HST observations have disclosed a shell-like structure of several
bright HII regions; Lynds et al. \cite{lynds} measured the shell's
expansion velocity of $50 - 70$~km/s, however, Lozinskaya et
al.~\cite{our_1} did not confirm these results. Silich et al.
\cite{silich} found the traces of a faint giant   ring sized
$D\simeq 500$\,pc in the H$_\alpha$ emission of diffuse gas.

In our paper \cite{our_1} we have investigated the structure and
kinematics of ionized gas in VII~Zw~403. In addition to the
previously known bright HII regions and traces of a faint giant
ring,  many new faint diffuse and arc-shaped structures were
found, and a ``fine structure'' of the giant ring was identified.
We found no evidence of the shell-like regions expanding with the
velocity of \mbox{50--70~km/s}, which was reported
in~\cite{lynds}, but instead we detected a clear line broadening
(FWHM up to \mbox{60--120\,km/s}) in the region of weak diffuse
emission outside the bright HII regions. In the brightest shell
¹~1 around the richest and youngest association ¹~1  faint
details are revealed in the wings of the [OIII] line at the
velocities of up to \mbox{$-200 \div -300$~km/s,} in the wings
of \mbox{H$_\alpha$} \mbox{line---at} the velocities of up to
$-350$~km/s, and up to \mbox{550--600~km/s} from the line center
\cite{our_1}. Such velocities in VII~Zw~403 were discovered for
the first time and are a clear evidence of gas acceleration at the
shock front. The kinematic age of the bright shells,
corresponding to our estimates of their average expansion velocities
(20~km/s or below) is at least 2--4 Myr, which agrees well with
the age of their central OB associations, reported
in~\cite{lynds}. Faint extended filamentary and diffuse regions of
ionized gas that can be observed in the entire central part of the
galaxy, as well as the giant HII ring can be related to the
older stellar population of the latest star formation burst (with
the age of 10~Myr according to~\cite{lynds}).

In \cite{our_2}, based on the observations of VII~Zw~403 at the
6-m BTA telescope of the SAO RAS with the  panoramic Multipupil
Fiber Spectrograph (MPFS)  and with the SCORPIO focal reducer  in
the
 slit spectrograph mode, we have analyzed the
gas emission spectrum and determined the relative oxygen abundance
in the regions ¹~1, 3, 4, and the nitrogen and sulfur abundances
in all regions. To our knowledge, these estimates of
metallicity in individual regions of ionized gas in VII~Zw~403
were the first, all the previously published values are averaged
across the galaxy.

The startup of the Spitzer Space Telescope has opened new avenues
for studying the dust component of the interstellar medium of
galaxies. The  polycyclic aromatic hydrocarbons (PAH): the macromolecules
 consisting of
tens of and hundreds of atoms, mostly hydrogen and carbon ~
\cite{PAH_ARAA}, are of
particular interest. The mechanism of formation and destruction of
PAH molecules is not yet completely understood, the discussed possibilities
are: their formation in the atmospheres of carbon-rich AGB and
post-AGB stars or in the dense molecular clouds, and the
destruction of these molecules by the shock waves or UV emission
of hot stars (see, e.g., Sandstrom et al.~\cite{sandstrom} and
references therein).

To clarify the mechanisms of formation and destruction of PAHs and
their relationship with the physical parameters and metallicity of
gas in the galaxies, Sandstrom et al. \cite{sandstrom}, based on
the Spitzer observations have investigated the   dust component of
the nearby dIrr galaxy Small Magellanic Cloud. Our own detailed
analysis of the infrared emission of  dIrr galaxy IC~10 has shown
that the results of observations are better consistent with the
assumption that the PAHs form in the molecular clouds and are
destroyed by the UV emission, rather than being formed in the
atmospheres of carbon-rich stars~\cite{wiebe}. To elucidate the
nature of PAHs, it may be important to refer to the observed
deficiency of radiation of these molecules in the galaxies with
low metallicity. According to Drain et al.\cite{draineetal2007},
this deficiency is related to the shortage of PAHs themselves,
rather than to the less efficient excitation of the IR transitions
in them. In our work \cite{wiebe}, based on the observations of
HII regions in the IC~10 galaxy, we for the first time suspected a
correlation between the mass fraction of dust contained in the
PAHs ($q_{\rm PAH}$), and metallicity of gas at the level of
individual HII regions. However, the accuracy of the discovered
dependence was found to be insufficient to consider this
correlation reliable, and we find it interesting to test it on the
observations of other galaxies.

In this paper we refine the abundances of oxygen, nitrogen and
sulfur in certain HII regions from the results of previous
observations of VII~Zw~403 with the \mbox{6-m}  BTA telescope of
the Special Astrophysical Observatory of Russian Academy of
Sciences (SAO RAS) \cite{our_2, our_1}. Based on the archive data
of the Spitzer Space Telescope  we investigate the infrared
emission of the galaxy with the aim to analyze the distribution of
the PAHs and find a possible correlation of $q_{\rm PAH}$ with the
radiation brightness of the ionized and neutral gas, as well as
with the metallicity of HII regions.

The following sections describe the observations we used and
present the obtained results, and the conclusion summarizes our
main findings.

The distance to VII~Zw~403 is adopted as 
\mbox{$D=4.5$\,Mpc} \cite{lynds, shul_lad_a}, which corresponds to
an angular scale of 22\,pc$/''$.

\section{OBSERVATIONS USED}

\subsection{Spectral Observations}

Spectral observations of VII~Zw~403 were performed at the 6-meter
SAO RAS BTA telescope with the SCORPIO focal reducer\footnote{{\tt
http://www.sao.ru/hq/lsfvo/devices/scorpio/ 
/scorpio.html}} \cite{scorpio} in the spectrograph slit  mode, as
well as  with the  panoramic Multipupil Fiber Spectrograph (MPFS).
Both series of observations, carried out in 2003 and 2004 are
described in detail in our paper \cite{our_2}, we shall only
briefly summarize the basic data here.

\begin{figure*}
\center{\vspace{3mm}\includegraphics[width=0.55\textwidth]{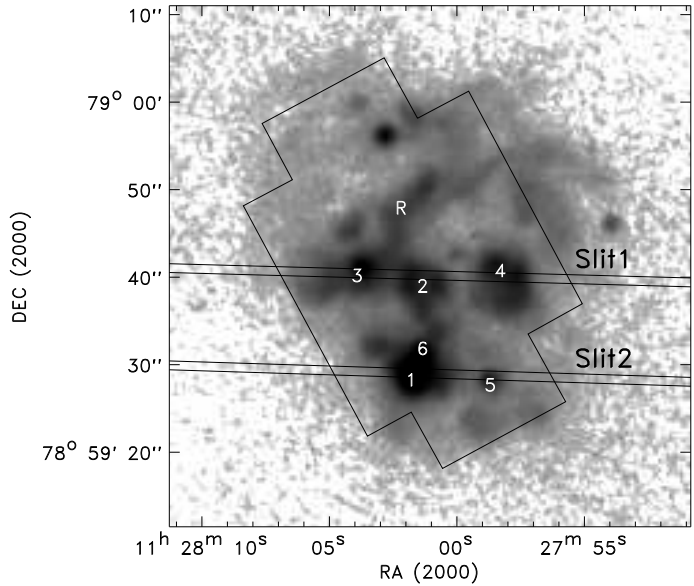}}
\caption{The region of the latest star formation burst
in VII~Zw~403 viewed in the H$_\alpha$ line. The localization of
the spectrograph slit and a part of the galaxy, overlapped with
the MPFS observations are shown. The numbers indicate the HII
complexes (according to the markings by Lynds et al. \cite{lynds}),
the letter R marks the brightest part of the giant ring.
}\label{fig_loc}
\end{figure*}

In the observations with a slit spectrograph, the slit length was
about $6'$,  its width was $1''$, the scale along the
slit---$0.36''$/ pixel. Two spectrograms  were obtained, the
spectrograph slit localization is shown in Fig.~\ref{fig_loc}.
Slit~1 passed through the complexes ¹~2--4, Slit~2---through
the complexes \mbox{¹~1 and 5.} For two slit positions we obtained
the spectra in the ranges of 6270--7300\,\AA\ and 4800--5600\,\AA,
the seeing was about $2''$, the exposure ranged from 800 to
1800\,s. The data reduction was done using the standard method,
the spectrophotometric standard AGK+81$^\circ$266 was applied to
bind to the energy scale. The emission line intensities were
determined using a single-component Gaussian approximation of
profiles.

In the observations with the MPFS\footnote{{\tt
http://www.sao.ru/hq/lsfvo/devices/mpfs/ 
/mpfs\_main.html}} \cite{mpfs} ~ the spectra from 256 spatial
elements were simultaneously registered, making up the matrix in
the picture plane sized $16\times16$ elements, where the angular
size of the pixel is $1''$. The spectra with the resolution of
about 8\,\AA\, were obtained in the range of
\mbox{4250--7200\,\AA\,} for seven fields in the central region of
the galaxy, taken with a mutual offset. The total time of
exposures for each field varied within \mbox{600--1800\,s} with
the average seeing of   \mbox{$1.7'' - 2''$.}

The observations were reduced using the software packages
developed in the SAO RAS Laboratory of the Spectroscopy and
Photometry of Extragalactic Objects, running under IDL. The
spectra GRW+70$^\circ$5824 were used as the spectrophotometric
standard. The result of data reduction is the data cube, in which
each pixel of the image sized $16''\times16''$ corresponds to the
spectrum of 2048 elements. After the initial reduction the data
cubes for all seven fields were aligned and stacked, so that the
size of the resulting mosaic amounted to $49''\times 31''$; the
corresponding region of the galaxy is shown in Fig.~\ref{fig_loc}.

\subsection{Archival Observational Data of the Spitzer Space
Telescope}

We used the archival observational data of the galaxy VII~Zw~403
with the IRAC and MIPS instruments obtained within the program
called the Starburst Activity in Nearby Galaxies\footnote{Spitzer
Proposal, ID 59 (2003)}. The data were downloaded from the archive
using the Spitzer Heritage Archive system\footnote{{\tt
http://sha.ipac.caltech.edu}}. The images of the galaxy were
obtained in seven wavelengths: $3.6\mu$m, $4.5\mu$m, $5.8\mu$m,
$8.0\mu$m from the observations with the IRAC instrument, and
$24\mu$m, $70\mu$m, $160\mu$m---with the MIPS instrument. 
The images were combined in mosaic using the MOPEX software\footnote{{\tt
http://ssc.spitzer.caltech.edu/dataanalysistools/ 
/tools/mopex/}}; the  images in the $70\mu$m and $160\mu$m bands
were processed using the GeRT software package\footnote{{\tt
http://ssc.spitzer.caltech.edu/dataanalysistools/ 
/tools/gert/}}. For the analysis of mosaic images we used our own
procedures written in IDL. In addition, in order to determine the
integral fluxes as the level of background emission in all
infrared bands, we used the mean intensity in the positions,
distant from the main star-forming regions of the galaxy. In our
case, this method is applicable owing to the   small angular size
of VII~Zw~403; the galaxy occupied a small part of the field in
the images.

\section{RESULTS}

Figure~\ref{fig_loc} demonstrates the image of the star-forming
complex of VII~Zw~403 in the H$_\alpha$ line, obtained from our
observations with the 6-m \mbox{telescope \cite{our_1}.} The
figure shows the localization of two slits and the mosaic of
fields, observed with the MPFS; the numbers denote the numbers of
HII regions based on the markings by Lynds et al.\cite{lynds}, the
letter R marks the above-mentioned gigantic ring.

Determining the fluxes in the emission lines, we averaged
individual  regions to increase the signal-to-noise ratio.

To refine the metallicity in HII regions of the galaxy we used the
line intensities, previously measured in~\cite{our_2}, as well as
the data adopted from \mbox{Lynds et al. \cite{lynds}}. In this
paper we additionally take into account the relative intensities
of   weaker lines: [SIII]$\lambda6312$\,\AA/H$_\alpha$ for the
regions ¹~1--5 from the observations with a slit spectrograph,
and [OIII]$\lambda4363$\AA/H$_\beta$ for the region ¹~1 from
the MPFS observations. All the relative line intensities used (not
corrected for the interstellar absorption) are listed in
Table~\ref{tab_fluxes}.

\begin{table*}

 \caption{Relative line intensities in the spectra of five bright HII regions}
 \label{tab_fluxes} {\scriptsize\begin{tabular}{c|c|c|c|c|c|c} \hline
 Lines                 &   ¹~1    &   ¹~2     &  ¹~3     &  ¹~4     &   ¹~5   & reference \\
\hline
[OII]3729/3727         &$1.39\pm0.11$&$1.43\pm0.14$&$1.50\pm0.08$&$1.37\pm0.10$&$1.80\pm0.23$            &\cite{lynds}\\

[OII](3727+29)/H$_\beta$&$0.59\pm0.02$&$2.26\pm0.11$&$1.18\pm0.03$&$1.71\pm0.06$&$2.20\pm0.12$            &\cite{lynds}\\

[OIII](5007)/H$_\beta$  &$3.89\pm0.01$&$1.65\pm0.03$&$2.68\pm0.01$&$2.59\pm0.03$&$1.28\pm0.03$             &\cite{lynds}\\

[OIII](5007)/H$_\beta$  &$3.43\pm0.05$&$1.72\pm0.02$&$2.85\pm0.03$&$2.78\pm0.03$&$1.69\pm0.01$            & *\\

[OIII](5007)/H$_\beta$  &$3.86\pm0.02$&$1.79\pm0.04$&$2.85\pm0.02$&$2.82\pm0.04$&$1.63\pm0.07$& **\\

[OIII](4363)/H$_\beta$  &$0.064\pm0.002$&   ---   &$0.049\pm0.006$&$0.063\pm0.009$&     ---      &\cite{lynds}\\

[OIII](4363)/H$_\beta$  &$0.067\pm0.009$& --- & --- & --- & ---  & ** \\

[OI](6300)/H$_\beta$&$0.011\pm0.003$&$0.04\pm0.01$&$0.044\pm0.005$& ---  &    ---    &*\\

[SIII](6312)/H$_\alpha$&$0.0066\pm0.0014$&$0.0073\pm0.0038$&$0.0133\pm0.0052$& $0.0087\pm0.0031$  &  $0.0072\pm0.0059$ &*\\

[SII](6717+31)/H$_\alpha$&$0.066\pm0.001$&$0.137\pm0.001$&$0.123\pm0.001$&$0.096\pm0.001$&$0.137\pm0.003$ & *\\

[SII](6717+31)/H$_\alpha$&$0.051\pm0.001$&$0.140\pm0.008$&$0.110\pm0.003$&$0.092\pm0.005$&$0.137\pm0.008$& **\\

[SII](6717)/H$_\alpha$ &$0.028\pm0.001$&$0.07\pm0.012$&$0.06\pm0.003$&$0.056\pm0.005$&$0.076\pm0.09$& **\\

[NII](6548+84)/H$_\alpha$&$0.024\pm0.001$&$0.032\pm0.001$&$0.024\pm0.001$&$0.025\pm0.003$&$0.038\pm0.002$ &*\\

[NII](6584)/H$_\alpha$   &$0.013\pm0.001$&$0.027\pm0.001$&$0.022\pm0.001$&$0.021\pm0.001$&$0.034\pm0.002$ &*\\

H$_\gamma$/H$_\beta$     &$0.44\pm0.01$& $0.50\pm0.01$&$0.42\pm0.01$&$0.47\pm0.01$&$0.51\pm0.02$&\cite{lynds}\\

H$_\beta$/H$_\alpha$     &$0.313\pm0.002$&$0.304\pm0.005$&$0.307\pm0.002$&$0.309\pm0.004$&$0.255\pm0.01$&**\\

\hline
\end{tabular}
}

\begin{tabular}{rcl}
*&---&our data from the slit spectra. \\
**&---&our data from the MPFS observations.\\
\end{tabular}

\end{table*}

\subsection{Metallicity of Ionized Gas in the Galaxy}

In \cite{our_2} we estimated the relative abundances  of oxygen,
nitrogen and sulfur based on the assumption that inside the HII
regions  these elements predominantly exist in the
 OIII, NII, SII states, respectively. The rationale for this
followed from the analysis of radial distribution of relative
intensities of the corresponding lines inside the brightest region
¹~1. In this paper, we estimate the metallicity of gas from the
line intensities listed in Table~\ref{tab_fluxes}, taking into
account  the other ionization stages  as well. We applied the
well-known methods for estimating the  oxygen, nitrogen and sulfur
abundances: the classic ``direct'' \mbox{$T_e$-method} (Izotov et
al. \cite{izotov_method}), and a number of empirical methods. The
results are presented in Table~\ref{tab_abu} with the indications
of the methods used.

\subsubsection{The $T_e$-Method}

In order to use the $T_e$-method (Izotov \mbox{et al.
\cite{izotov_method})} we have to know the temperature $T_e$ and
density $n_e$ of gas (but the latter has little effect
on the result). The density is found from the lines of
[SII]~$\lambda6717$/$\lambda6731$; the temperature was estimated
with the {\tt Five Level} program \cite{5lev} from the line
intensity ratio of
[OIII]~($\lambda$4959\,\AA+$\lambda$5007\,\AA)/$\lambda$4363\,\AA
 lines. Our observations allowed to confidently measure
the intensity of the  [OIII]~$\lambda$4363\,\AA\, line only for
region ¹~1; the $T_e$ value we got coincides with the estimate
by \mbox{Lynds et al. \cite{lynds}}.

To determine the temperature in the HII regions ¹~3 and 4, we
used the observations from \cite{lynds}; while for the regions
¹~2 and 5, where the line [OIII]~$\lambda4363$\,\AA\, was
never measured, we adopted the temperature \mbox{$T_e=14500$,}
measured for \mbox{region ¹~1.}

Knowing $T_e$ and $ n_e $, we determined the abundance of ions
$\mathrm{O^+, O^{2+}}$ using the relations \mbox{(3) and (5)}
\mbox{from \cite{izotov_method}.} The full oxygen abundance is
found in the assumption that the oxygen in the HII regions
predominantly exists in the OII and OIII states. The abundances of
the nitrogen and sulfur ions are found from the relations (6), (8)
and (9) from [20]. To get the full abundance of these elements, we
calculated the ionization correction factors ($ICF$) for the unobserved
ionization stages from the relations (18) and (20) for the mean
metallicity of \mbox{$12+\log \mathrm{(O/H)}= 7.6$.} To determine
the abundances of $\mathrm{O^+, N^{+}, S^+}$ we adopted $T_e$ in
the emission region of these ions, computed from relation (14),
and to determine the $\mathrm{S^{2+}}$ abundance, the $T_e$  was
computed using relation (15) from \cite{izotov_method}.

 The estimates of the relative abundances of oxygen, nitrogen and sulfur
by the $T_e$-method are presented in Table~\ref{tab_abu} and
indicated by the index ``$T_e$''; the metallicity measurement
errors listed do not take into account the temperature estimation
error.

\subsubsection{Empirical methods}

A series of methods is currently widely used to estimate the
abundances of chemical elements in the HII regions. These methods
apply the empirical relationships between the metallicity and
relative line intensities in the spectrum. Their accuracy is
generally inferior to that of the  ``direct'' $T_e$-method.

However, since we determined  $T_e$ from our observations only for
the region ¹~1, we used the data from \cite{lynds} for the
regions ¹~3 and 4, and for regions ¹~2 and 5 there are no
estimates of $T_e$ at all, we  hence found it useful to apply the
empirical methods for estimating their metallicity.


We used the long-known empirical P-method (Pilyugin,
Thuan~\cite{pilyugin_method}), based on the dependence of oxygen
abundance  on the excitation parameter $P$ and the parameter
$R_{23}$, which are determined from the relative intensities of
the [OII] and [OIII] lines. As a result, we found that the
estimates computed by this technique and using the $T_e$-method
for different HII regions are in good agreement.


An  empirical dependence proposed by Pettini and Pagel
\cite{pettini_method} has an advantage, which is based on the
relationship of bright and closely spaced lines ([OIII]/H$_\beta$
and [NII]/H$_\alpha$), i.e. it is insensitive to absorption.
However, the values of the  O3N2 parameter, used in this
dependence, turned out to be outside the range of confident method
applicability for three regions of VII~Zw~403, and at its boundary
for two other regions.


We as well used nine empirical relations for various line
intensity ratios, proposed in \cite{nagao}, but they gave a large
(approximately $\pm0.5$\,dex) scatter of estimates for the same
region in different lines. At that, the mean metallicity estimates
based on different dependences are consistent with the estimates
of the $T_e$-method.


Pilyugin et al. \cite{pil_2010}, and Pilyugin and Mattson
\cite{pil_mat} have proposed the most accurate empirical methods
for estimating the metallicity. The accuracy of estimates,
obtained applying two techniques from the former work, is,
according to the authors, 0.075\,dex for the oxygen abundance and
0.05\,dex for the nitrogen abundance; the accuracy of methods
proposed in the latter work is evaluated as 0.077\,dex and
0.110\,dex for oxygen and nitrogen, respectively.

The results obtained for the HII regions of VII~Zw~403 based on
the empirical dependences from these papers are presented in
Table\,2.

The technique from \cite{pil_2010} allows us to estimate the
relative oxygen and nitrogen abundances from the ratio of
intensities of the following bright lines: \\
\mbox{([OIII]~$\lambda4959+5007$\,\AA)\slash H$_\beta$,}
 \mbox{([OII]~$\lambda3727+3729$\,\AA)\slash H$_\beta$,}
 \mbox{([NII]~$\lambda6548+6584$\,\AA)\slash H$_\beta$}. \linebreak
 The authors designate this method as \mbox{ON-method.}
When applying the ONS-method, another ratio
([SII]~$\lambda6717+6731$\AA)\slash H$_\beta$ is as well in use.
We have applied both methods; to estimate the relative abundances
of oxygen and nitrogen via the ONS-method we used the ratios (17)
and (18)  \mbox{from \cite{pil_2010};} via the ON-method---the
ratios (19) and (20) were applied. The results obtained are listed
in Table\,2, indicated by the indices ``ONS'' and ``ON'',
respectively.


The NS-method presented in \cite{pil_mat} possesses a certain
advantage for us as it does not require the measurement of
intensities of \mbox{([OII]~$\lambda3727+3729$\,\AA) lines.}
Applying this method, from the relative intensities of the
following bright lines: \\
\mbox{([OIII]~$\lambda4959+5007$\AA)\slash H$_\beta$,} 
\mbox{([NII]~$\lambda6548+6584$\AA)\slash H$_\beta$, and }
 \mbox{([SII]~$\lambda6717+6731$\AA)\slash H$_\beta$,}
we have found
 the relative oxygen abundance from relation
(8), and relative nitrogen abundance from relation (9) \mbox{in
\cite{pil_mat}.} The results are listed in Table\,2 and indicated
by the ``NS'' indices.

\begin{table*}

 \caption{Metallicity of HII regions}\label{tab_abu}
\center{\footnotesize\begin{tabular}{c|c|c|c|c|c} \hline
\multirow{2}{*}{Estimate} & \multicolumn{5}{c}{HII regions} \\
                  \cline{2-6}
              & ¹1 & ¹2 & ¹3 & ¹4 & ¹5 \\
\hline
\multicolumn{1}{l|}{$N_e$(SII) cm$^{-3}$} &   57    &   66    &     57 &       110&    --   \\
\multicolumn{1}{l|}{$T_e$(OIII) K} &  14500  &    --   &     14900&     16200&     -- \\
\hline

\multicolumn{1}{l|}{$12+\log\mathrm{(O/H)}$ ($T_e$)}&$ 7.69 \pm 0.01 $&$ 7.69 \pm 0.03 $&$ 7.66 \pm 0.01$&$ 7.64 \pm 0.02 $&$ 7.68 \pm 0.03 $\\

\multicolumn{1}{l|}{$12+\log\mathrm{(O/H)}$ (ONS)} & $7.72 \pm 0.02$ & $7.75 \pm 0.03$ & $7.68 \pm 0.02$ & $7.71 \pm 0.03$ & $7.79 \pm 0.04$ \\
\multicolumn{1}{l|}{$12+\log\mathrm{(O/H)}$ (ON)} &  $7.80 \pm 0.02$ & $7.55 \pm 0.02$ & $7.61 \pm 0.02$ & $7.65 \pm 0.04$ & $7.59 \pm 0.03$ \\
\multicolumn{1}{l|}{$12+\log\mathrm{(O/H)}$ (NS)} &  $7.77 \pm 0.01$ & $7.56 \pm 0.01$ & $7.62 \pm 0.01$ & $7.67 \pm 0.03$ & $7.60 \pm 0.02$ \\

\hline

\multicolumn{1}{l|}{$12+\log\mathrm{(N^+/H)}$}&$5.72 \pm 0.02 $&$ 5.86 \pm 0.02 $&$ 5.72 \pm 0.02$&$ 5.70 \pm 0.05 $&$ 6.01 \pm 0.03 $\\

\multicolumn{1}{l|}{$ICF\mathrm{(N^+)}$}&$6.82 \pm 0.22 $&$ 1.75 \pm 0.10 $&$ 3.27 \pm 0.08$&$ 2.45 \pm 0.09 $&$ 1.70 \pm 0.11  $\\

\multicolumn{1}{l|}{$12+\log\mathrm{(N/H)}$ ($T_e$)}&$6.55 \pm 0.02 $&$ 6.10 \pm 0.03 $&$ 6.23 \pm 0.02$&$ 6.09 \pm 0.05 $&$ 6.24 \pm 0.04 $\\

\multicolumn{1}{l|}{$12+\log\mathrm{(N/H)}$ (ONS)} &  $6.48 \pm 0.03$ & $6.13 \pm 0.04$ & $6.09 \pm 0.03$ & $6.11 \pm 0.07$ & $6.22 \pm 0.06$ \\
\multicolumn{1}{l|}{$12+\log\mathrm{(N/H)}$ (ON)} &  $6.56 \pm 0.02$ & $6.01 \pm 0.03$ & $6.09 \pm 0.02$ & $6.09 \pm 0.06$ & $6.09 \pm 0.04$ \\
\multicolumn{1}{l|}{$12+\log\mathrm{(N/H)}$ (NS)} &  $6.32 \pm 0.02$ & $6.07 \pm 0.02$ & $6.09 \pm 0.02$ & $6.18 \pm 0.04$ & $6.13 \pm 0.03$ \\

\multicolumn{1}{l|}{$12+\log\mathrm{(S^+/H)}$}&$5.39 \pm 0.01 $&$ 5.73 \pm 0.03 $&$ 5.60 \pm 0.01$&$ 5.50 \pm 0.02 $&$ 5.79 \pm 0.03 $\\

\multicolumn{1}{l|}{$12+\log\mathrm{(S^{2+}/H)}$}&$6.07 \pm 0.09 $&$ 6.12 \pm 0.23 $&$ 6.23 \pm 0.16$&$ 5.96 \pm 0.15 $&$ 6.12 \pm 0.35$\\

\multicolumn{1}{l|}{$ICF\mathrm{(S^{+}+S^{2+})}$}&$1.32 \pm 0.02 $&$ 1.05 \pm 0.01 $&$ 1.10 \pm 0.01$&$ 1.06 \pm 0.01 $&$ 1.05 \pm 0.01 $\\

\multicolumn{1}{l|}{$12+\log\mathrm{(S/H)}$}&$6.27 \pm 0.07 $&$ 6.29 \pm 0.16 $&$ 6.36 \pm 0.13$&$ 6.11 \pm 0.11 $&$ 6.31 \pm 0.24 $\\

\hline
\end{tabular}}
\end{table*}


We  further use the metallicity estimates of individual HII
regions, listed in Table\,~\ref{tab_abu}, obtained by the direct
$T_e$-method and by the most accurate empirical methods
\cite{pil_2010, pil_mat}, which are yielding results in good
agreement with the $T_e$-method.

In particular, for regions ¹~1, 3, 4 we adopted the estimates
of the $T_e$-method, while for regions ¹~2, 5, in which there
are no $T_e$ measurements, we averaged the listed values that were
determined with different techniques.

As we can see from Table\,\ref{tab_abu}, the metallicity in
individual HII regions does not reveal strong variations and is
well consistent with the mean values over the galaxy, obtained by
other authors. According to our estimate, the average metallicity
in VII~Zw~403 amounts to
\mbox{$12+\log(\mathrm{O/H})=7.66\pm0.03$;} according to Izotov
and al.~\cite{izot_obs} it is equal to 
$12+\log(\mathrm{O/H}) = 7.73 \pm0.01$; Nagao et al. \cite{nagao}
used the line intensities measured in \cite{izot_obs}  to obtain
 \mbox{$12+\log(\mathrm{O/H}) = 7.7\pm0.01$;} while
Schulte-Ladbeck et al. \cite{shul_lad_b} give
\mbox{$12+\log\mathrm{(O/H)}=7.63 - 7.71$.}

The abundance of nitrogen in different regions is also consistent
with the average across the galaxy: 
$12+\log(\mathrm{N/H}) = 6.19\pm0.08$ according to
\cite{izot_obs}.

Our estimates of the sulfur abundance are higher than the average
across the galaxy:   \mbox{$12+\log(\mathrm{S/H}) =
6.16\pm0.04$} according to \cite{izot_obs}, but given the
measurement errors they are consistent with this value for all
regions, except region ¹~3.

The nitrogen abundance, averaged across the galaxy, according to
our evaluation is  \mbox{$12+\log(\mathrm{N/H}) =
6.22\pm0.19$,} the sulfur abundance amounts to
\mbox{$12+\log(\mathrm{S/H}) = 6.27\pm0.14$.} Note that for
VII~Zw~403 we found practically equal abundances of nitrogen and
sulfur in the HII regions. Similar abundances of these elements
are observed in other galaxies, such as the Magellanic
\mbox{Clouds \cite{garnett99}} and the NGC~3109 \cite{pena}. The
galaxy NGC~3109 is the closest analogue of VII~Zw~403 based on the
abundance of O, N and S.

\subsection{The Dust and Polycyclic Aromatic Hydrocarbons in VII~Zw~403}

Our particular interest to the IR observations of VII~Zw~403 stems
from the fact that our analysis of the dust component of the Irr
galaxy IC~10 resulted in shedding some light on the as  yet
unclear mechanisms of the formation and destruction of the PAH
molecules \cite{wiebe}. We will hence attempt to similarly
consider the IR observations of VII~Zw~403.

\begin{figure*}

 \center{\includegraphics[width=0.8\textwidth]{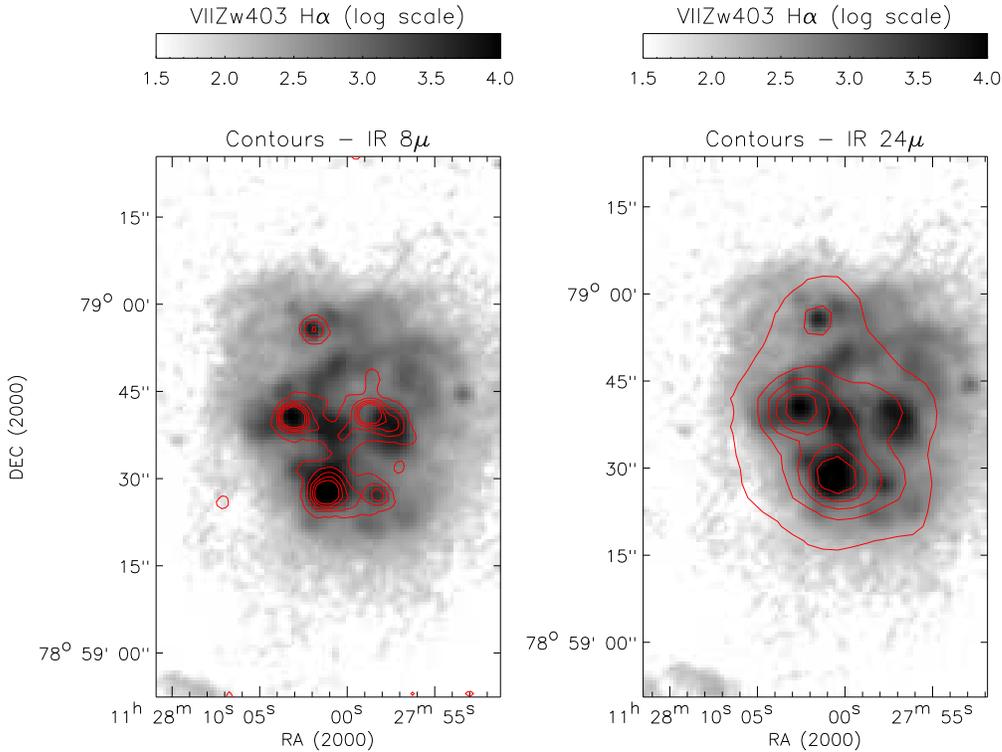}}
\caption{An image of the star-forming region of VII~Zw~403 in
the H$_\alpha$ line with the superimposed contours in different IR
range bands: left---8~$\mu$m, right---24~$\mu$m. The isophotes,
corresponding to the levels of 0.1, 0.2, 0.3, 0.4 and 0.5 MJy/sr
for 8~$\mu$m and 0.2, 0.9, 1.6, 2.3, 3.0 MJy/sr for ~24~$\mu$m are
shown.} \label{fig_ir}
\end{figure*}

Figure~\ref {fig_ir} demonstrates the maps of the infrared
emission distribution in VII~Zw~403 in the 8~$\mu$m and 24~$\mu$m
bands, constructed from
the Spitzer observational data,  combined with the \mbox{H$_\alpha$ image}.

The photometry of VII~Zw~403 in different IR bands from the
Spitzer observations  was earlier conducted by Engelbracht et
al.~\cite{engel}. We have performed the subsequent aperture
photometry of the entire galaxy based on the same data, the
results of which are listed Table~\ref{tab_ir_flux}. Our estimates
somewhat vary from the results in \cite{engel}, the differences
are most  probably due to different values of the adopted
background level in the images; however, these differences have
practically no influence on the results discussed below.

In the analysis of the infrared radiation Draine and ~Li
\cite{dl2007} proposed to parameterize the UV field of the galaxy
as the sum of the ``minimal'' diffuse UV field   $U_{\min}$,
filling the  largest part of its volume, and the more intensive UV
field with a power law energy distribution, affecting only the
mass fraction   $\gamma $ of all the dust in the galaxy. $U_{\min}$, expressed in the units of average UV-field of our Galaxy, characterizes the overall level of star formation in
the studied system, and  $\gamma$ allows to estimate
what fraction of the galaxy's matter is involved in the current
processes of star formation. Draine and ~\mbox{Li \cite{dl2007}}
give the algorithm for estimating the parameters of the galaxy
from the observations in the infrared bands of 8~$\mu$m,
24~$\mu$m, 70~$\mu$m and 160~$\mu$m (the data for the 3.6~$\mu$m
band is used to remove the contribution of the stellar radiation).
We used the results of photometry of  VII~Zw~403 to apply the
methods proposed \mbox{in \cite{dl2007}} and try to determine
$U_{\min}$, $\gamma$, and \mbox{$q_{\rm PAH}$,} or the mass
fraction of dust contained in the PAHs. Unfortunately, the values
of the parameters found for VII~Zw~403, that appear in the models
from \cite{dl2007}, are beyond the limits, for which these models
are calculated. Therefore, we could not determine the $U_{\min}$
and $\gamma$, we only got the upper limit of
$q_\mathrm{PAH}<0.5\%$ and the lower limit of $U_{\min}>25$. This
value of $U_{\min}$ indicates a high level of star formation in
VII~Zw~403, which is natural for BCD galaxies. A low abundance of
PAHs is consistent with the hypothesis of the destruction of these
molecules under the effect of UV radiation in the regions of star
formation near the hot stars.

A number of authors, in particular Sandstrom et al.
\cite{sandstrom} note the possibility of using the flux ratio at
the wavelengths of 8~$\mu$m and 24~$\mu$m as a local indicator of
the mass fraction of $q_{\rm PAH}$. Based on the correlation that
the above authors have found between the $q_{\rm PAH}$ and the
ratio $F_{8\mu {\rm m}}/F_{24\mu {\rm m}}$, we have analyzed the
distribution of $q_{\rm PAH}$ in VII~Zw~403 and tried to compare
it with the distribution of ionized and neutral gas in the galaxy.

\begin{figure*}
 \center{\includegraphics[width=0.9\textwidth]{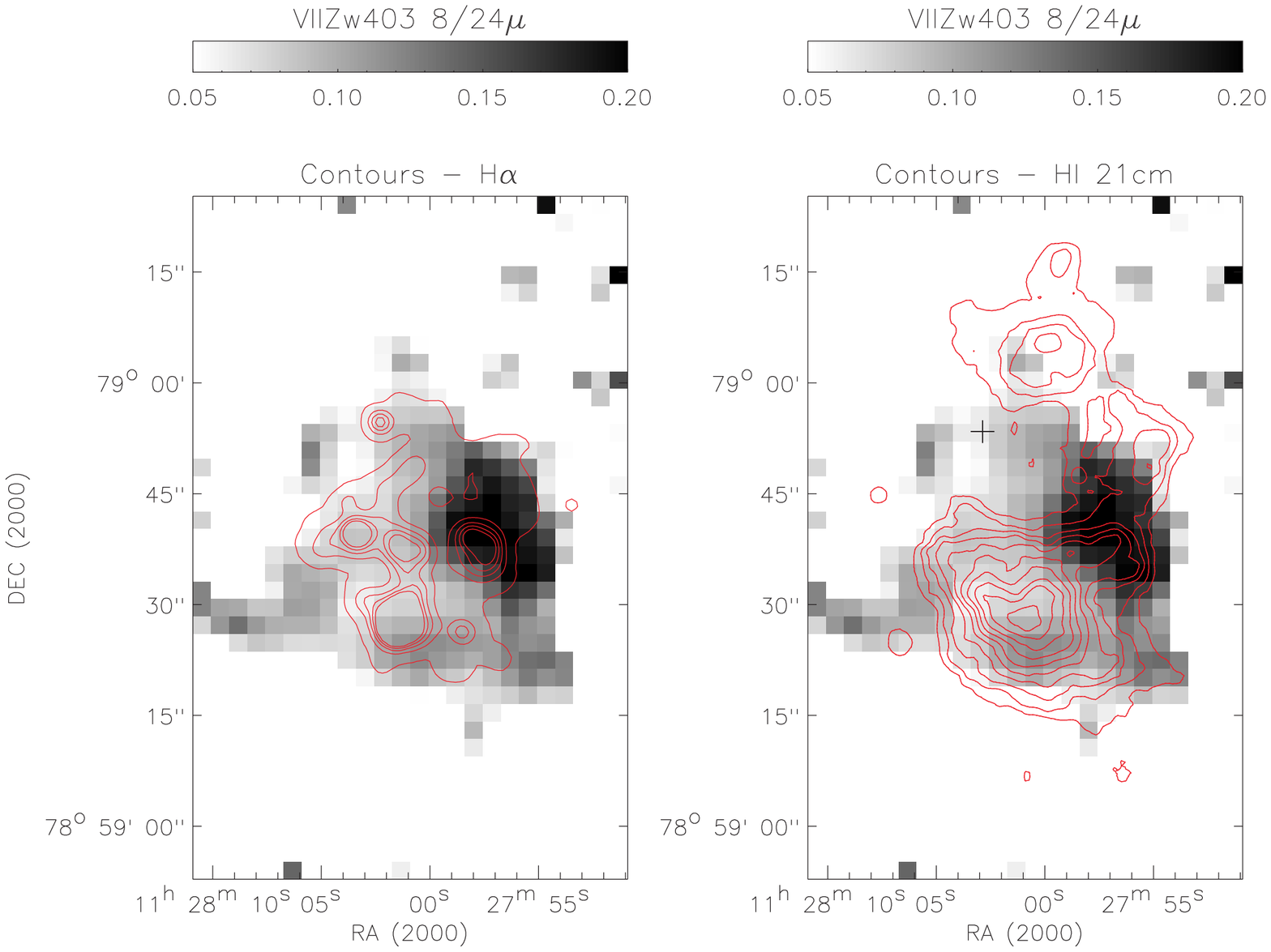}}
 \caption{The maps of the flux  ratio $F_{8\mu {\rm m}}/F_{24\mu {\rm m}}$
 of the VII~Zw~403  star-forming region with the
superimposed contours in H$_\alpha$ (left) and HI 21\,cm (right).
The isophotes in the H$_\alpha$ line correspond to the following
brightness levels: 500, 1750, 3000, 4250, 5500 in arbitrary units;
the isophotes in the HI 21~cm line indicate the brightness levels,
corresponding to the beam density of N(H)=(5, 10, 15, 20, 25, 30,
35, 40, 45 and 50)$\times10^{20}$ atoms/cm$^2$. The cross on the
right-hand side plot marks the position of a single point X-ray
source detected in this galaxy  \cite{ott_point}. }
\label{fig_824}
\end{figure*}

Figure~\ref{fig_824} demonstrates the maps of distribution of the
flux ratio  $F_{8\mu {\rm m}}/F_{24\mu {\rm m}}$  with the
superimposed isophotes in the lines of H$_\alpha$ (left) and HI
21\,cm (right). We used the isophotes in the 21\,cm line from the
VLA observations, presented by Ashley and  Simpson
\cite{ashley}.

\begin{figure*}
\center{\includegraphics[width=0.6\textwidth]{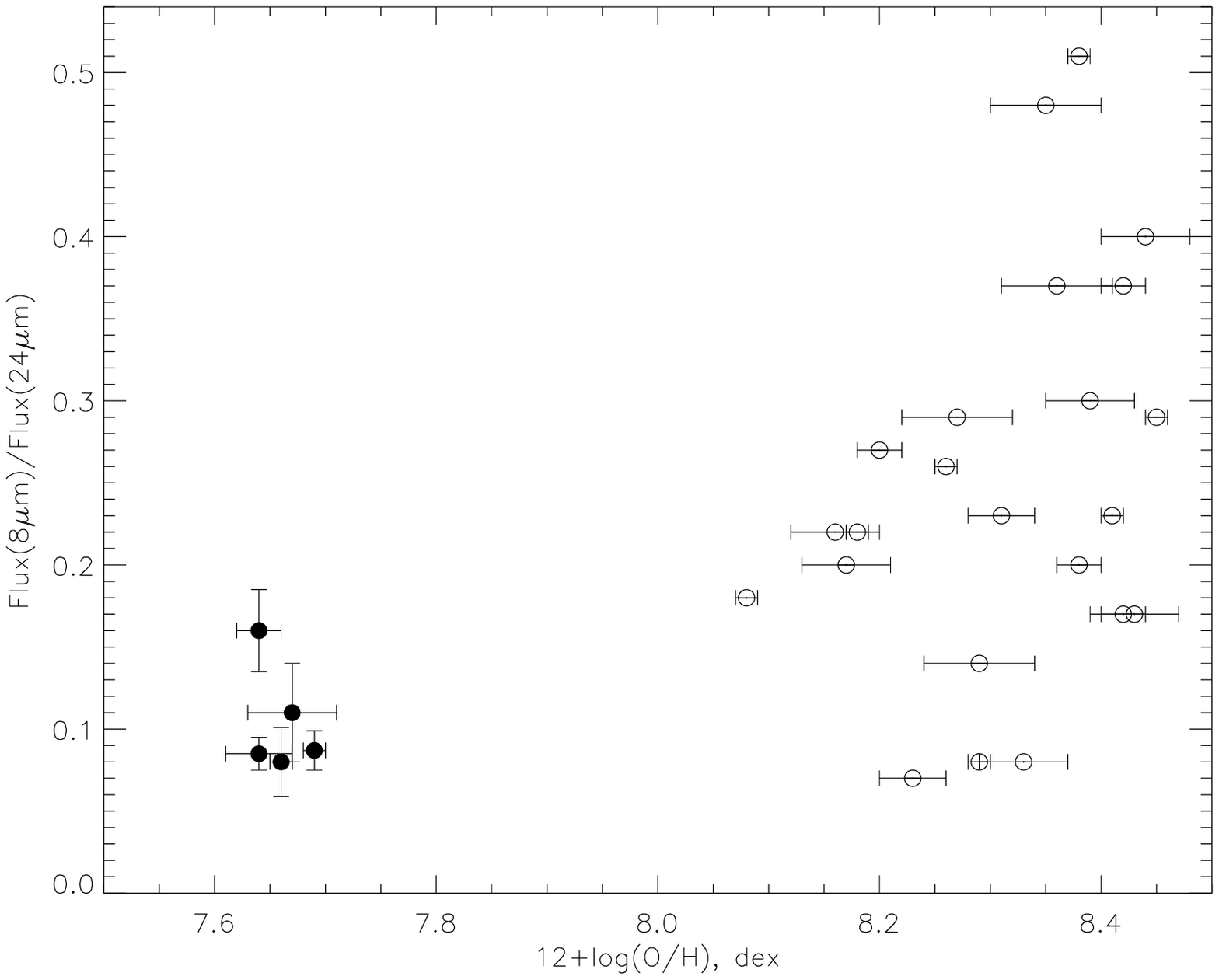}}
\caption{The flux ratio $F_{8\mu {\rm m}}/F_{24\mu {\rm m}}$
depending on the oxygen abundance for  individual HII regions
based on the results of study of two galaxies---VII~Zw~403 (filled
circles, according to the present work) and IC~10 (empty circles,
according to \cite{wiebe}). The flux ratio errors  for the HII
regions in IC~10 do not exceed 0.04.} \label{fig_ir_met}
\end{figure*}

It follows from Fig.~\ref{fig_824} that  large variations of the
flux ratio $F_{8\mu {\rm m}}/F_{24\mu {\rm m}}$ are not revealed
in the central part of VII~Zw~403, except for the bright western
region. However, we clearly see that this ratio is increased at
the borders and outside the HII regions, and declines in their
central parts. This may be due to the destruction of the PAH
molecules in the bright regions of  ionized gas under the effect
of ultraviolet radiation of the central OB associations.

\begin{table*}[tbp]

\caption{The results of photometry of the images from the Spitzer
Space Telescope}\label{tab_ir_flux}
\begin{tabular}{c|c|c|c|c|c|c}
\hline
$F_{3.6\mu {\rm m}}$, mJy & $F_{4.5\mu {\rm m}}$, mJy & $F_{5.8\mu {\rm m}}$, mJy & $F_{8.0\mu {\rm m}}$, mJy & $F_{24\mu {\rm m}}$, mJy & $F_{70\mu {\rm m}}$, Jy & $F_{160\mu {\rm m}}$, Jy \\
\hline
$3.27\pm0.02$ & $1.94\pm0.02$ & $2.02\pm0.06$ & $3.58\pm0.07$ & $28.4\pm0.7$ & $ 0.355\pm0.003$ & $0.123\pm0.003$\\
\hline
\end{tabular}
\end{table*}

Due to the low spatial resolution we could not discern in
VII~Zw~403  a  distinct correlation of the PAH distribution with
the envelopes of neutral hydrogen, similar to that observed in
IC~10 \cite{wiebe}. As follows from Fig.\,3, the highest value of
$F_{8\mu {\rm m}}/F_{24\mu {\rm m}}$, corresponding to the largest
mass fraction of PAH, occurs in the direction of the western part
of the dense HI cloud. It is possible that in the direction of the
southern part of the cloud, the PAH molecules are destroyed by the
ultraviolet emission of the OB associations located here. Violent
ultraviolet emission is as well revealed by the brightest in
galaxy shell region ¹~1 located here, see
Fig.~\ref{fig_loc}.


The galaxies of low metallicity are revealing a deficiency of PAH
radiation, linked with a shortage of these macromolecules (see
Drain et al.\cite{draineetal2007} and references therein). In the
galaxies with the oxygen abundance of  $12+\log({\rm O/H}) > 8.1$
the typical values  are   $q_{\rm PAH}\sim 4\%$, in the galaxies
with lower metallicity, the average value corresponds to $q_{\rm
PAH}\simeq 1\%$. In our \mbox{work \cite{wiebe}} from the
observations of IC~10 we have suspected the existence of a similar
correlation between the PAH mass fraction and the metallicity of
gas at the level of individual HII regions within one galaxy, and
not only in the comparison of different galaxies. We hypothesized
that this correlation may reflect the processes of the PAH
molecule formation, but not their subsequent evolution. However,
the accuracy of the dependence, observed in IC~10 proved to be
insufficient to consider this correlation credible. Hence, it is
interesting to test whether in VII~Zw~403 there occurs a drop in
$q_{\rm PAH}$ with the decreasing metallicity of  HII regions.

Figure~\ref{fig_ir_met} shows the flux ratio  $F_{8\mu {\rm
m}}/F_{24\mu {\rm m}}$ for the individual  HII zones as a function
of metallicity for the two galaxies studied: IC~10 according to
Wiebe et al.~\cite{wiebe} and VII~Zw~403 from the results of this
work.

Note that in~\cite{wiebe} we used the metallicity value, obtained
with the O3N2-method \cite{pettini_method}. For consistency, we
tried to estimate the relative oxygen abundance in the  HII
regions of the IC~10  galaxy with the
\mbox{NS-method~\cite{pil_mat},} used above for VII~Zw~403.
However, the low intensity of the [NII]~$\lambda6548$\,\AA\ and
[OIII]~$\lambda4959$\,\AA\  lines  in the weak HII regions has
increased the measurement errors, although the general form of the
dependence has not changed significantly. This is why in
Fig.~\ref{fig_ir_met} for IC~10 we list the results obtained with
the O3N2-method according to the data from~\cite{wiebe}.

In \cite{wiebe}, we noted that the ratio $F_{8\mu {\rm
m}}/F_{24\mu {\rm m}}$ apparently drops with decreasing oxygen
abundance in the HII regions of IC~10, but the threshold value is
not \mbox{$12+\log({\rm O/H}) = 8.0-8.1$,}  as observed comparing
different galaxies, but rather about $8.3$. Unfortunately, small
variations of metallicity and the ratio $F_{8\mu {\rm m}}/F_{24\mu
{\rm m}}$ in the HII regions of VII~Zw~403 do not allow to confirm
from the observations of this galaxy the presence of the
correlation, earlier suspected in IC~10. Such a dependence is only
discernible if we consider both galaxies together, but its
existence at the level of a comparison between different galaxies
has been already known before.

\section{CONCLUSION}

We have analyzed the archival infrared observations of the BCD
galaxy VII~Zw~403 from the Spitzer Space Telescope and re-analyzed
the results of our observations at the 6-m BTA telescope of the
SAO RAS with a slit spectrograph and the MPFS, published in
\cite{our_2}. In \cite{our_2} our estimations of the oxygen,
nitrogen and sulfur abundances were based on the assumption that
the gas in the bright HII regions is located primarily in the
OIII, NII, and  SII states. In the present work, we additionally
took into account the other stages of ionization, used weaker
lines, and made the new metallicity estimates applying several
different methods.

Note that the  metallicity measurements in individual regions of
ionized gas in VII~Zw~403 were never previously attempted, except
in our earlier work \cite{our_2}.

The oxygen abundance of the individual HII regions and the average
value across the galaxy  found in the present study and amounting
to  \mbox{$12+\log(\mathrm{O/H})=7.66\pm0.03$} is fully
consistent with the estimates of the average metallicity in
 VII~Zw~403 determined by other authors: 
\mbox{$12+\log(\mathrm{O/H}) = 7.73\pm0.01$} according \mbox{to
\cite{izot_obs};} Nagao et al. \cite{nagao} from the measurements
of \cite{izot_obs} have obtained \mbox{$12+\log(\mathrm{O/H}) =
7.7\pm0.01$;} according  to \cite{shul_lad_b}
\mbox{$Z=0.05-0.06 Z_\odot$,} which corresponds to 
$12+\log\mathrm{(O/H)}=7.63 - 7.71$ taken the solar metallicity,
adopted in this paper \mbox{$12+\log\mathrm{(O/H)_\odot}=8.93$.}

One of the tasks of the iterative analysis of spectroscopic
observations, as well as the analysis of infrared images from the
Spitzer Space Telescope  was the search for the correlation of the
PAH mass fraction with the metallicity at the level of individual
HII regions that we suspected in the study of the IC~10 galaxy
\cite{wiebe}.

However, the variations of metallicity and the ratio $F_{8\mu {\rm
m}}/F_{24\mu {\rm m}}$ in the HII regions of VII~Zw~403, which can
serve as an indicator of the PAH mass fraction, were small, which
did not allow us to confirm the existence of such a correlation
within this galaxy.

The analysis of the images in the infrared range has yielded an
estimate of the PAH mass fraction averaged across the galaxy
amounting to \mbox{$q_{PAH}<0.5\%$,} what is indicative of the
vigorous star formation going on in VII~Zw~403. A comparison of
the distribution of $F_{8\mu {\rm m}}/F_{24\mu {\rm m}}$ in the
galaxy with an image in the H$_\alpha$ line reveals an elevated
PAH abundance at the boundaries and beyond the bright HII regions,
and, respectively, a decreased PAH abundance in their inner
regions. A similar effect is also observed in other galaxies, in
particular in the Small Magellanic  Cloud
\cite{sandstrom} and IC~10 ~\cite{wiebe}.

A comparison of the map of $F_{8\mu {\rm
m}}/F_{24\mu {\rm m}}$ with the emission in the HI  21\,cm line
has not demonstrated any distinct correlation of the PAH abundance
with the distribution of neutral gas in the VII~Zw~403 galaxy.
Nonetheless, we note that the maximum value of $F_{8\mu {\rm
m}}/F_{24\mu {\rm m}}$ is observed towards the western part of the
giant dense HI clouds. Perhaps a lower value of $F_{8\mu {\rm
m}}/F_{24\mu {\rm m}}$ in the other parts of this cloud is caused
by the destruction of PAHs under the influence of ionizing
radiation of the OB associations located there.

The patterns revealed are consistent with the assumption that the
PAH molecules form in the giant molecular clouds, to be later
destroyed by the ultraviolet radiation.

\begin{acknowledgments}
The work was conducted with the financial support from the RFBR
grant (project ¹~10-02-00091). O.~V.~Egorov is grateful for the
financial support of the non-profit Dmitry Zimin's Dynasty
Foundation and the Federal Target Programm Scientific and
Scientific-Pedagogical Cadre of Innovative Russia (state contract
14.740.11.0800). The authors thank the anonymous referee for the
helpful comments and V.~P.~Arkhipova and D.~S.~Wiebe for the
stimulating discussions. The study is based on the observational
data obtained at the 6-m SAO RAS telescope, founded by the Russian
Ministry of Science (registration ¹~01-43). We made use of the
NASA/IPAC Extragalactic Database (NED),   operated by the Jet
Propulsion Laboratory on the campus of the California Institute of
Technology, under contract with NASA (USA).

\end{acknowledgments}

\flushright{\textit{Translated by A.~Zyazeva}}

\end{document}